\documentclass[namedreferences]{solarphysics}
\usepackage[optionalrh]{spr-sola-addons} 
\usepackage{graphicx}
\usepackage{color}
\usepackage{url}             

\newcommand{\adv}{    {\it Adv. Space Res.}}

\newcommand{\aap}{    {\it Astron. Astrophys.}}

\newcommand{\apj}{    {\it Astrophys. J.}}
\newcommand{\apjl}{   {\it Astrophys. J. Lett.}}

\newcommand{\grl}{    {\it Geophys. Res. Lett.}}

\newcommand{\jgr}{    {\it J. Geophys. Res.}}

\newcommand{\pasj}{   {\it Publ. Astron. Soc. Japan}}

\newcommand{\solphys}{{\it Solar Phys.}}

\newcommand{\ssr}{    {\it Space Sci. Rev.}}

\ifx \doiurl    \undefined \def
\doiurl#1{\href{http://dx.doi.org/#1}{\textsf{DOI}}}\fi \ifx
\adsurl \undefined \def
\adsurl#1{\href{http://adsabs.harvard.edu/abs/#1}{\textsf{ADS}}}\fi
\ifx \arxivurl  \undefined \def
\arxivurl#1{\href{http://arxiv.org/abs/#1}{\textsf{arXiv}}}\fi

\begin{document}
\begin{article}
\begin{opening}

\title{Development and Parameters of a Non-Self-Similar CME Caused by Eruption of a Quiescent Prominence}

\author{\inits{I.V.}\fnm{I.V.}~\lnm{Kuzmenko}$^{1}$\orcid{0000-0001-5119-5723}} \sep
\author{\inits{V.V.}\fnm{V.V.}~\lnm{Grechnev}$^{2}$\orcid{0000-0001-5308-6336}}

 \runningauthor{I.V. Kuzmenko, V.V. Grechnev}
 \runningtitle{Development of a Non-Flare-Related CME}

\institute{$^{1}$ Ussuriysk Astrophysical Observatory, Solnechnaya
                  St. 21, Primorsky Krai, Gornotaezhnoe 692533, Russia
                  email: \url{kuzmenko_irina@mail.ru} \\
            $^{2}$ Institute of Solar-Terrestrial Physics SB RAS,
                  Lermontov St.\ 126A, Irkutsk 664033, Russia
                  email: \url{grechnev@iszf.irk.ru}}

\date{Received ; accepted }

\begin{abstract}
The eruption of a large quiescent prominence on 17 August 2013 and
associated coronal mass ejection (CME) were observed from different
vantage points by \textit{Solar Dynamics Observatory} (SDO),
\textit{Solar-Terrestrial Relations Observatory} (STEREO), and
\textit{Solar and Heliospheric Observatory} (SOHO). Screening of the
quiet Sun by the prominence produced an isolated negative microwave
burst. We estimated parameters of the erupting prominence from a
model of radio absorption and measured from 304\,\AA\ images. Their
variations obtained by both methods are similar and agree within a
factor of two. The CME development was studied from the kinematics
of the front and different components of the core and their
structural changes. The results are verified using movies in which
the CME expansion was compensated according to the measured
kinematics. We found that the CME mass ($3.6 \times 10^{15}$\,g) was
mainly supplied by the prominence ($\approx 6 \times 10^{15}$\,g),
while a considerable part drained back. The mass of the
coronal-temperature component did not exceed $10^{15}$\,g. The CME
was initiated by the erupting prominence, which constituted its core
and remained active. The structural and kinematical changes started
in the core and propagated outward. The CME structures continued to
form during expansion, which did not become self-similar up to
$25\,\mathrm{R}_\odot$. The aerodynamic drag was insignificant. The
core formed until $4\,\mathrm{R}_\odot$. Some of its components were
observed to straighten and stretch forward, indicating the
transformation of tangled structures of the core into a simpler flux
rope, which grew and filled the cavity as the CME expanded.
\end{abstract}


\end{opening}

\section{Introduction}
\label{S-introduction}

Prominence eruptions can be associated with most significant
manifestations of solar activity such as coronal mass ejections
(CMEs) and flares. Clouds of magnetized plasma hitting Earth are
able to cause hazardous space-weather disturbances. Solar eruptions
have been known for many years; nevertheless, their scenarios,
responsible processes, and parameters of erupted magnetized plasma
still need clarification. In spite of a large body of observational
material supplied by modern solar telescopes, the existing concepts
are mainly based on traditional hypotheses proposed several decades
ago and near-Earth \textit{in-situ} measurements extrapolated to the
Sun.

The main problems preventing considerable progress in understanding
solar eruptions are caused by difficulties in their observations and
measuring their parameters. One of the causes is the low brightness
of erupting structures, which rapidly fade during expansion
concurrently with increasing flare emission. Next, it is not
possible to observe the CME development in a single spectral range
starting from its genesis up to distances of several solar radii
($\mathrm{R}_\odot$), which makes difficult identification of the
structures visible by different instruments. Furthermore, it is only
possible to estimate physical characteristics of the eruptions and
CMEs by means of indirect methods, while the object of the
measurements is poorly defined, and its properties are not known
exactly.

According to the modern view, the main active structure of a CME is
a magnetic flux rope (MFR), which governs its development and
subsequent expansion. Some researchers assume an MFR to pre-exist
before the eruption onset \citep{Chen1989, Chen1996, Cheng2013}.
Some others relate the MFR formation to reconnection processes also
responsible for solar flares \citep{InhesterBirnHesse1992,
LongcopeBeveridge2007, Qiu2007}. There are different views on the
kinematics of the erupting structures and CMEs that reflect the
forces governing their expansion. Reviews of the existing problems,
observations, and scenarios under discussion have been given by
\cite{Gopalswamy2004} and \cite{Forbes2006} (see also
\citealp{Grechnev2015}). The MFR is mainly considered as a rather
uniform magnetic structure identified with the CME cavity. According
to the traditional view, the MFR is enclosed in a turbulent sheath,
and its bottom part contains a frozen-in dense core that inherits
the material of the prominence, whose role in the CME genesis is
passive.

The CME development and formation is traditionally associated with a
flare in an active region or with a prominence eruption outside of
active regions occurring without pronounced flare manifestations.
CMEs of both types are probably caused by processes that are
basically similar but have different quantitative parameters; some
qualitative dissimilarity has also been found (\textit{e.g.}
\citealp{ChertokGrechnevUralov2009}). An additional category of CMEs
that are not accompanied by any detectable surface activity has been
identified in the last decade \citep{Robbrecht2009}. While
flare-related eruptions have been extensively studied in recent
years, lesser attention has been paid to non-flare-related eruptions
of ``quiescent'' prominences outside of active regions.

Eruptions of prominences (filaments) are observed in different
spectral ranges such as the visible light (the H$\alpha$ line), in
extreme-ultraviolet (EUV, the best-suited is the He\,{\sc ii}
304\,\AA\ line), and in microwaves. A filament eruption is sometimes
accompanied by a ``negative burst'', \textit{i.e.} a temporary
decrease of the total microwave flux below a quasi-stationary level.
Such phenomena were discovered by \cite{CovingtonDodson1953}, who
interpreted them as absorption of radio emission in material of an
erupting prominence. Later studies confirmed this idea and led to a
scenario of screening a microwave source by a cloud of
low-temperature absorbing material \citep{Covington1973,
Sawyer1977}. The dependence of the absorption depth on both the
radio frequency and properties of absorbing plasma makes it possible
to estimate some parameters of the responsible erupting structure,
if a microwave depression is observed at different frequencies.
Thus, negative bursts can provide information about eruptions.

This consideration motivated our studies of several events with
negative bursts \citep{KuzmenkoGrechnevUralov2009, Grechnev2011,
Grechnev2013}. Negative bursts are rarely observed and usually
follow an ordinary flare-related impulsive burst. The time-profiles
and depression depths are dissimilar at different frequencies. To
reproduce this behavior, we developed a model calculating absorption
at different radio frequencies in a screen of given dimensions,
temperature, and density, assuming a simple flat-layered geometry of
the screen \citep{Grechnev2008, KuzmenkoGrechnevUralov2009}.
Modeling absorption of the total microwave flux observed at
different frequencies provided estimates of the absorbing material
even without images. Studies of combined data observed in different
ranges of solar emission show that a typical cause of depressions is
screening of both a compact microwave source and large areas of the
quiet Sun. Almost all of the events analyzed were associated with
flares in active regions, when erupted prominence material screened
a radio source located in the same or a nearby active region. Rare
cases of negative bursts preceding an impulsive burst or lacking it
have been studied insufficiently. We are not aware of events in
which only quiet-Sun regions were screened.

These studies used mainly the observations in the past, whose
opportunities were considerably poorer than now. An imaging interval
as long as six hours was typical of observations in the 304\,\AA\
channel, in which eruptive prominences are best visible. The current
observational opportunities are considerably broadened due to the
\textit{Atmospheric Imaging Assembly} (AIA: \citealp{Lemen2012AIA})
onboard the \textit{Solar Dynamics Observatory} (SDO). The situation
is still more favorable, when the Sun is additionally observed from
different vantage points by the \textit{Sun-Earth-Connection Coronal
and Heliospheric Investigation} instrument suite (SECCHI:
\citealp{Howard2008}) onboard the \textit{Solar-Terrestrial
Relations Observatory} (STEREO: \citealp{Kaiser2008}).

In this article we study the eruption of a quiescent prominence away
from active regions on 16\,--\,17 August 2013, which caused an
isolated negative burst without any impulsive burst or a flare.
Total-flux microwave data of a satisfactory quality are available at
several frequencies. The high imaging rate of SDO/AIA in the
304\,\AA\ channel allows comparison of the model estimates from
radio data at several times with evolving parameters of the eruptive
prominence directly measured from the images.

The sets of EUV and white-light images available make it possible to
follow the appearance of the CME near the Sun and its expansion up
to distances exceeding $20\,\mathrm{R}_\odot$. One of the main
methods to study CMEs is based on the measurements of their
structural components. The most important characteristic is
acceleration, which reflects the dynamics of acting forces. However,
acceleration is the second derivative of measurable characteristics,
and its calculation by means of differentiation leads to
considerable uncertainties. Invoking the standard methods to
estimate the measurement errors might not be adequate here, because
the main uncertainty lies in the identification of the feature in
question and is unknown.

To overcome these difficulties, we use a different approach based on
an analytic fit of a smooth function to the experimental
measurements \citep{Gallagher2003, Sheeley2007, WangZhangShen2009}.
A bell-shaped acceleration corresponds to the fact that the initial
and final velocities of an eruption are nearly constant. A
particular shape of the acceleration is insignificant, because a
double integration is required to reproduce the measurable
distance--time points. This approach was justified in preceding
studies (\textit{e.g.} \citealp{Grechnev2015, Grechnev2016}).

Pursuing reliability of the kinematic measurements, we endeavor to
reveal possible changes in the CME shape and structure around
presumable acceleration episodes. To facilitate their comparison at
different times, we compensate for the CME expansion by resizing the
images according to the measured kinematics, in which the CME
appears static \citep{Grechnev2014, Grechnev2015, Grechnev2016}.
This method appears to be the most appropriate so far to assess the
measurement accuracy. The conclusion whether a structure in question
is static or not is easily drawn from the visual inspection of a
movie. It is more difficult to assess the measurement quality from a
usual set of non-resized images by means of any image-processing
method (\textit{e.g.} \citealp{Maricic2004, Bein2011}), because the
CME structures appear nonuniform and progressively fade in the
images.

Section~\ref{S-description} briefly describes the event. In
Section~\ref{S-estimates} we estimate parameters of erupted plasma
from microwave data and compare them with the measurements from
the EUV images. Section~\ref{S-kinematics} is devoted to the
kinematics of the eruptive prominence becoming the CME core as
well as the frontal structure from overlapping images of different
spectral ranges. The results are discussed in
Section~\ref{S-discussion} and summarized in
Section~\ref{S-conclusion}.


\section{Description of the Event}
 \label{S-description}

The eruption of a large quiescent prominence was observed by SDO/AIA
in 304\,\AA\ starting at about 22:50 on 16 August 2013 (all times
hereafter refer to UTC). To study the event, we used data from
several online data centers. The SDO/AIA level~1.5
quarter-resolution data with an interval of two to four minutes were
taken from \url{jsoc.stanford.edu/data/aia/synoptic/}. The
STEREO/EUVI images with a ten-minute interval are available at
\url{sharpp.nrl.navy.mil/cgi-bin/swdbi/secchi_flight/img_short/form}.
We used microwave total-flux data recorded by the \textit{Nobeyama
Radio Polarimeters} (NoRP: \citealp{Torii1979, Nakajima1985};
\url{ftp://solar.nro.nao.ac.jp/pub/norp/xdr/}), the US Air Force
\textit{Radio Solar Telescope Network} (RSTN:
\url{ftp://ftp.ngdc.noaa.gov/STP/space-weather/solar-data/solar-features/solar-radio/rstn-1-second/}),
and the Ussuriysk Observatory Radiometer at 2.8~GHz (RT-2:
\citealp{Kuzmenko2008}; \url{www.uafo.ru/observ_rus.php}, station
code VORO).

The lists and movies of CMEs as well as their parameters measured
from the images produced by the \textit{Large Angle and
Spectroscopic Coronagraph} (LASCO: \citealp{Brueckner1995}) onboard
SOHO are available in the online CME catalog (\citealp{Yashiro2004};
\url{cdaw.gsfc.nasa.gov/CME_list/}). The images produced by the C2
and C3 LASCO coronagraphs with an interval of 12 minutes were taken
from \url{sohowww.nascom.nasa.gov/data/archive.html}. We also used
the images produced by the STEREO-B coronagraphs: COR1 with
intervals of five to ten minutes and COR2 with intervals of
15\,--\,30 minutes
(\url{sharpp.nrl.navy.mil/cgi-bin/swdbi/secchi_flight/img_short/form}).

\begin{figure} 
   \centerline{\includegraphics[width=\textwidth]
    {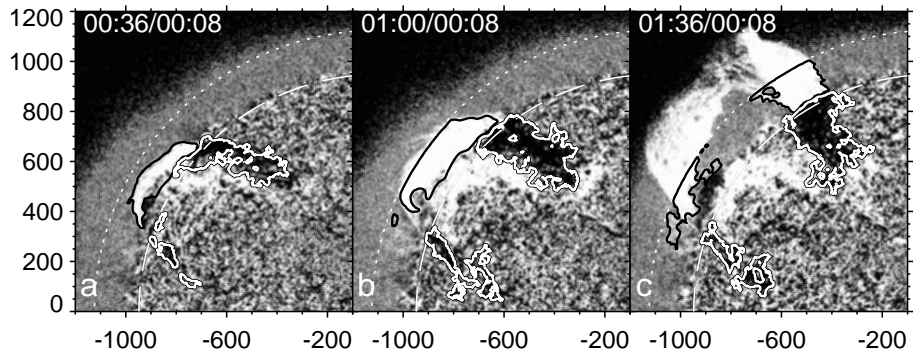}
   }
   \caption{Three episodes of the prominence eruption in
SDO/AIA 304\,\AA\ image ratios. The \textit{white-dotted circle}
corresponds to the solar radio radius at 1\,GHz
($1.186\,\mathrm{R}_\odot$). The portions of the prominence
considered in estimations are outlined by the \textit{white contour}
on the solar disk and by the \textit{black contour} above the limb.
The axes indicate the distance from solar disk center in
arcseconds.}
   \label{F-aia304}
\end{figure}

The rising prominence was visible until, at least, 02:00 on 17
August, and its south leg is detectable after 03:00. The AIA
304\,\AA\ image ratios in Figure~\ref{F-aia304} present the
prominence, which was located in the North-East quadrant of the Sun
away from activity complexes. The prominence appears dark on the
solar disk because of absorption of the background solar emission in
its material. A large bright crescent on the disk is a negative
appearance of a pre-eruptive prominence visible in the base image at
00:08. Expansion of the rising prominence is manifested in large
dark patches moving on the solar disk, while the prominence is
bright above the limb. Its top part near the north leg loses opacity
in Figure~\ref{F-aia304}c.

The erupting prominence was also observed from STEREO-B spacecraft
located $138^{\circ}$ behind the Earth
(\url{cdaw.gsfc.nasa.gov/stereo/daily_movies/2013/08/17/}).
STEREO-A produced only one 304\,\AA\ image in two hours. We
therefore use STEREO-B data in this study. The
\textsf{20130817\_EUVI304.mpg} movie in the supplementary material
presents the prominence eruption observed by STEREO-B/EUVI in
304\,\AA. The contrast of the images was enhanced by dividing them
by an azimuthally-averaged radial background distribution. The
bases of the prominence were behind the limb for STEREO-B. A
bright region on the disk was not related to the eruption. The
movie reveals a complex threadlike structure of the prominence,
its untwisting, and draining cool plasma from its body. The top
part of the prominence near its north leg seems to stretch ahead.
Further details are discussed in Section~\ref{S-kinematics}.

According to the LASCO CME catalog,
starting from 01:26, SOHO/LASCO coronagraphs observed a weakly
accelerating CME with a central position angle of $42^{\circ}$,
which corresponds to the orientation of the erupting prominence. The
CME had an estimated mass of $3.6 \times 10^{15}$\,g, average speed
of 369\,km\,s$^{-1}$, and average acceleration of 5.1\,m\,s$^{-2}$.
Noticeable is a possible reacceleration of the CME at a distance
from the Sun around $20\,\mathrm{R}_\odot$, suggested by height--time
measurements in the catalog. The CME was also observed by the
coronagraphs on STEREO-B and STEREO-A. The CME is visible in the
\textsf{20130817\_cor1\_orig.mpg} movie composed from the
STEREO-B/COR1 images in the polarized brightness, which reveal CMEs
without subtraction. The CME had a classical three-part structure
with a faint frontal structure (FS), cavity behind it, and a bright
core in the bottom part of the CME. The core corresponded to the
erupting prominence.

\begin{figure} 
   \centerline{\includegraphics[width=0.5\textwidth]
    {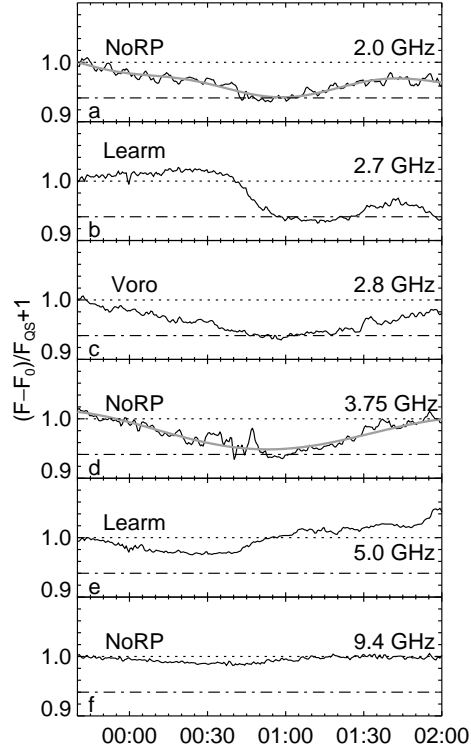}
   }
   \caption{Total-flux temporal profiles of the negative burst
at different frequencies normalized to the corresponding levels of
the quiet-Sun emission [$F_\mathrm{QS}$]. The pre-burst level
[$F_0$] at each frequency is subtracted.}
   \label{F-neg_burst}
\end{figure}

According to soft X-ray GOES-15 data, a weak B5.5 flare occurred
around 01:30 in an active region located at S21\,W56, far away from
the eruption region, being therefore irrelevant. Neither Type II or
Type III radio bursts nor an ``EUV wave'' accompanied the prominence
eruption. In microwaves, a negative burst corresponding to the
eruptive event was recorded at Nobeyama, Ussuriysk, and Learmonth.
Figure~\ref{F-neg_burst} presents total flux time-profiles of radio
emission at different frequencies. The pre-burst flux levels
[$F_\mathrm{b}$] are subtracted, and the data are smoothed with a
boxcar corresponding to 60 seconds and normalized to the quiet Sun
level [$F_\mathrm{QS}$] at each frequency. The NoRP data at 2 and
3.75\,GHz with considerable variations were fitted with a polynomial
(the gray thick line in Figures \ref{F-neg_burst}a and
\ref{F-neg_burst}d) for their subsequent processing. Unlike a
typical situation, the negative burst was ``isolated'', not being
preceded by the usual flare-related impulsive burst. At all
frequencies, except for 2.7\,GHz, the total flux started decreasing
below a quasi-stationary level at about 23:40 on 16 August. The
maximum depth reached $\approx 6.5\,\%$ of the quiet-Sun level at
01:00 on 17 August in a range of 2\,--\,3.75\,GHz, and then a
gradual recovery started. The quasi-stationary level at 5\,GHz and
9.4\,GHz recovered earlier than at lower frequencies. The depression
at 1\,GHz was neither deep nor long.

\section{Parameters of the Erupting Prominence}
  \label{S-estimates}

Screening of large quiet-Sun areas by the absorbing material of an
erupting filament can considerably contribute to the microwave
depression in a negative burst \citep{KuzmenkoGrechnevUralov2009,
Grechnev2011, Grechnev2013}. In the 16\,--\,17 August 2013 event, no
active regions existed on the path of the erupting prominence.
Hence, no compact radio sources could be screened. The only possible
cause of the negative burst was absorption of the emission from the
parts of the quiet Sun covered by the erupting prominence. From the
total-flux data available at a number of frequencies, parameters of
the erupting prominence can be estimated by means of a simple slab
model of an absorbing cloud.

\subsection{Model of Radio Absorption}

\begin{figure} 
   \centerline{\includegraphics[width=0.7\textwidth]
    {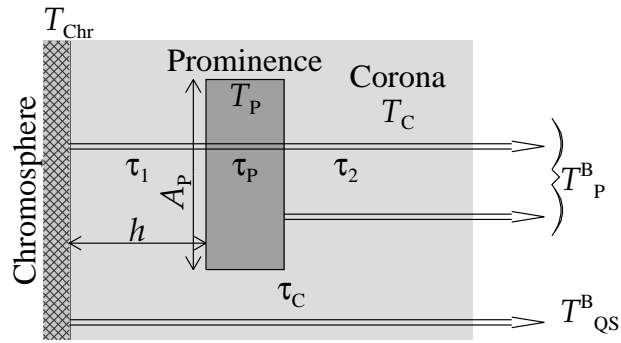}
   }
   \caption{A model of radio absorption used to estimate parameters
of the erupting prominence from observations of a negative
burst.}
   \label{F-abs_model}
\end{figure}

The model \citep{Grechnev2008, KuzmenkoGrechnevUralov2009}
considers the absorbing cloud as a uniform slab ``inserted'' into
the corona at some height [$h$] above the chromosphere
(Figure~\ref{F-abs_model}) and calculates the brightness
temperature after each layer as a sum of its own emission and a
non-absorbed remaining emission from preceding layers.

The model contains (i)~the chromosphere,
(ii)~the prominence of an area $A_\mathrm{P}$, kinetic temperature
$T_\mathrm{P}$, and optical thickness $\tau_\mathrm{P}$ at a height
[$h$] above the chromosphere, (iii)~a coronal layer between the
chromosphere and prominence of an optical thickness $\tau_1$, and
(iv)~a coronal layer between the prominence and observer of an
optical thickness $\tau_2$. The temperature of the corona is
$T_\mathrm{C} \approx 1.5 \times 10^6$\,K and that of the
chromosphere is $T_\mathrm{Chr} \approx 10^4$\,K. The total flux of
a negative burst [$F$] to the quiet-Sun total flux $F_\mathrm{QS}$
ratio is
\begin{eqnarray}
    F/F_\mathrm{QS} =
    [T^{\mathrm{B}}_\mathrm{QS}(A_\odot-A_\mathrm{P})+T^{\mathrm{B}}_\mathrm{P}
    A_\mathrm{P}]/(T^{\mathrm{B}}_\mathrm{QS}A_\odot). \nonumber
    \label{flux_ratio_eqn}
\end{eqnarray}
\noindent Here $T^{\mathrm{B}}_\mathrm{QS}$ and
$T^{\mathrm{B}}_\mathrm{P}$ are the brightness temperatures of the
quiet Sun and prominence, $A_\odot(\nu)$ and $A_\mathrm{P}$ are
the areas of the solar disk and the prominence. The brightness
temperature of the prominence is
\begin{eqnarray} T^{\mathrm{B}}_\mathrm{P} = T_\mathrm{Chr}e^{-(\tau_1 + \tau_2 +
        \tau_\mathrm{P})}
    + T_\mathrm{C}(1-e^{-\tau_1})e^{-(\tau_2+\tau_\mathrm{P})}
    \label{model_eqn} \nonumber \\
    +\ T_\mathrm{P}(1-e^{-\tau_\mathrm{P}})e^{-\tau_2} +
    T_\mathrm{C}(1-e^{-\tau_2}). \nonumber
\end{eqnarray}
\noindent Here $\tau_2=\tau_\mathrm{C}\exp(-2h/H)$, $H =
2kT_\mathrm{C}/(m_i g_\odot) \approx 8.4 \times 10^9\,
\mathrm{cm}$ is the height of the uniform atmosphere, $g_\odot =
274$\,m\,s$^{-2}$ solar gravity acceleration at the photosphere,
$\tau_1 = \tau_\mathrm{C} - \tau_2$, and $\tau_\mathrm{C}$ is
calculated from an equation $T^{\mathrm{B}}_\mathrm{QS} \approx
T_\mathrm{Chr}+T_\mathrm{C} \tau_\mathrm{C}$. The quiet-Sun
brightness temperature and radio radius at each frequency are
interpolated from reference values measured by \cite{Borovik1994}.
To keep the model self-consistent, we have used in the
calculations the reference brightness temperature and radio
radius, and the fluxes were calculated from these values.

The input parameters of the model are the optical thickness
[$\tau_\mathrm{P}$] of the absorbing cloud at a fiducial frequency
of 17\,GHz, its kinetic temperature [$T_\mathrm{P}$], area
[$A_\mathrm{P}$], and a height [$h$] of its lower edge above the
chromosphere. Adjusting the four parameters, we endeavor to reach
best fit of the total-flux spectrum computed from the model with
the absorption depths actually observed at different frequencies.

\subsection{Estimated Parameters}
 \label{S-estimated_parameters}

Parameters of erupting filaments were previously estimated from
radio absorption for the deepest depression or/and for the
observation time of a single 304\,\AA\ image, if it was available
\citep{Grechnev2008, Grechnev2011, Grechnev2013,
KuzmenkoGrechnevUralov2009}. Detailed SDO/AIA 304\,\AA\ data on
this event allow us to compare direct observations with the
temporal variations of the parameters estimated from radio
absorption. The 2.7\,GHz data were not used because of their
questionable stability. The results of the estimates from the
model are listed in Table~\ref{T-model_est}. The temperature of
the absorbing material of $\approx 9000$\,K did not change, the
optical thickness at 17\,GHz decreased from 0.7 to 0.01, the
height of the cloud increased from 100\,Mm to $\approx 200$\,Mm,
and the area increased from 3\,\% to $\approx 10\,\%$ of the
visible solar disk area [$A_\odot$] in an interval from 00:00 to
01:30. The estimate for each parameter was obtained by its
sequential least-squares optimizing. The errors listed in
Table~\ref{T-model_est} characterize the quality of the model fit
to the actual radio absorption spectrum.
Variation of the parameters within these
error ranges does not change significantly the sum of the squared
deviations between the fit and measurements.

\begin{table} 
 \caption{Parameters of the erupting prominence estimated from the model of radio absorption.}
 \label{T-model_est}
 \begin{tabular}{ccrcc}

 \hline

Time [UTC] & $\tau _\mathrm{17\,GHz}$ & \multicolumn{1}{c}{$A/A_\odot$ [\%]} & $h$ [Mm] & $T$ [MK] \\

 \hline

 00:00 & $0.70\pm 0.10 $ & $3.1 \pm 0.1$  &  $110 \pm 10$  & $9000 \pm 500$   \\
 00:10 & $0.70\pm 0.10 $ & $4.2 \pm 0.1$  &  $110 \pm 10$  & $9000 \pm 500$   \\
 00:20 & $0.70\pm 0.10 $ & $5.2 \pm 0.1$  &  $110 \pm 10$  & $9000 \pm 500$   \\
 00:30 & $0.60\pm 0.10 $ & $6.2 \pm 0.1$  &  $110 \pm 10$  & $9000 \pm 500$   \\
 00:40 & $0.30\pm 0.10 $ & $8.6 \pm 0.2$  &  $130 \pm 10$  & $9000 \pm 500$   \\
 00:50 & $0.09\pm 0.01 $ & $9.5 \pm 0.2$  &  $130 \pm 10$  & $9000 \pm 500$   \\
 01:00 & $0.06\pm 0.01 $ & $10.5 \pm 0.1$  &  $160 \pm 10$  & $9000 \pm 500$   \\
 01:10 & $0.035\pm 0.005 $ & $10.5 \pm 0.2$  &  $170 \pm 20$  & $9000 \pm 500$   \\
 01:20 & $0.03\pm 0.002 $ & $10.2 \pm 0.1$  &  $190 \pm 40$  & $9000 \pm 500$   \\
 01:30 & $0.01\pm 0.001 $ & $9.9 \pm 0.1$  &  $210 \pm 50$  & $9000 \pm 500$   \\

 \hline

 \end{tabular}

 \end{table}

On the other hand, the images in the 304\,\AA\ channel allowed us to
estimate the height of the prominence above the limb from
STEREO-B/EUVI data and its area from SDO/AIA data. Absorption of
radio emission is only possible when the solar disk is screened by
the prominence. When the prominence exits off-limb, the absorption
disappears. To get comparable estimates, we limited the area of the
prominence in the 304\,\AA\ images by a disk with a radius of
$1.186\,\mathrm{R}_\odot$ corresponding to the solar radio radius at
the lowest frequency of 1\,GHz, at which the negative burst was
observed. The area considered in the measurements is limited in
Figure~\ref{F-aia304} by the white contour on the disk (at a 15\,\%
brightness decrease) and by the black contour above the limb (at a
10\,\% brightness increase).

\begin{figure} 
   \centerline{\includegraphics[width=0.6\textwidth]
    {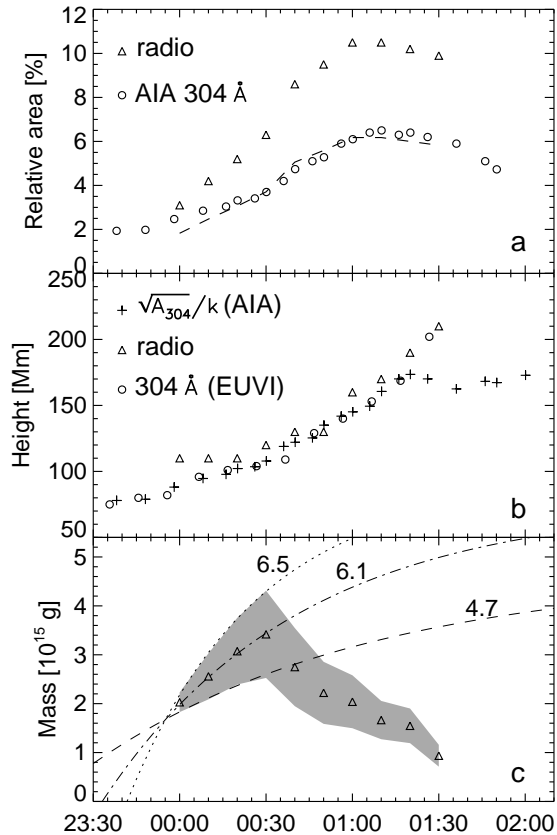}
   }
   \caption{Parameters of the erupting prominence measured
from the AIA 304\,\AA\ images and estimated from radio absorption
within a radius $1.186\,\mathrm{R}_\odot$. (\textbf{a})~Percentage
of the solar-disk coverage. The \textit{dashed line} represents the
area estimated from radio absorption divided by a factor of 1.7.
(\textbf{b})~The height of the lower edge estimated from radio
absorption (\textit{triangles}), measured from STEREO-B/EUVI
304\,\AA\ images (\textit{circles}), and estimated from the area
measured from SDO/AIA 304\,\AA\ images using a model shown in
Figure~\ref{F-model} (\textit{crosses}). (\textbf{c})~The estimated
mass of the erupted material (\textit{triangles}). The
\textit{shading} represents the uncertainties.}
   \label{F-prom_estimates}
\end{figure}

Figure~\ref{F-prom_estimates}a presents the variations of the
prominence area (percentage of the optical-disk area) measured from
the 304\,\AA\ images (circles) and those estimated from radio
absorption (triangles). The overall temporal behaviors of the two
data sets are similar to each other. Both sets represent an increase
of the projected part of the solar surface covered by the expanding
prominence until 01:05\,--\,01:20. Then the area decreases, because
the prominence loses opacity and departs from the analyzed region.
The temporal difference between the maxima estimated from radio and
EUV data is within the measurement errors.

The values estimated from radio absorption systematically exceed the
measurements from the EUV data. Comparison of the two sets is
facilitated by the dashed line in Figure~\ref{F-prom_estimates}a,
which represents the area estimated from radio absorption divided by
a factor of 1.7. The prominence area computed from the 304\,\AA\
images within the contours shown in Figure~\ref{F-aia304} might be
underestimated, because the contours are sensitive to the contrast
of the image, as their complex shapes indicate. Unlike this
situation, the estimates from radio absorption depend on an integral
effect, irrespective of the thickness of the absorbing layer. On the
other hand, the disadvantages of our model can result in an
overestimated area. The geometry assumed in the model, with layers
normal to the line of sight, is acceptable near the solar disk
center, but it strongly differs from the situation present near the
limb. Furthermore, the model does not consider the
frequency-dependent center-to-limb variation of the brightness
temperature. With the complications listed, the quantitative
difference between the estimates of the prominence area from radio
and EUV data within a factor of two appears to be acceptable, while
the two methods present almost the same temporal variations.

\begin{figure} 
   \centerline{\includegraphics[width=0.4\textwidth]
    {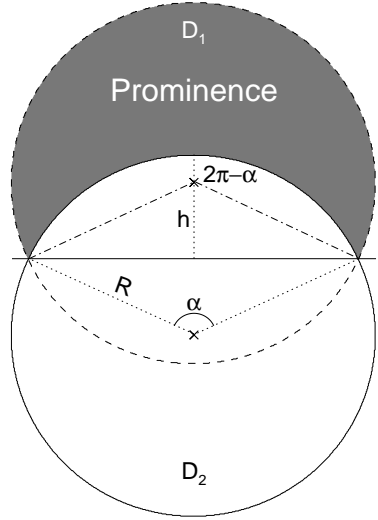}
   }
   \caption{A simple geometric model relating the shaded area
[$A$] of a crescent prominence with its height [$h$].}
   \label{F-model}
\end{figure}

We also estimated from radio absorption and measured the height of
the lower prominence edge above the photosphere from the 304\,\AA\
images. The height was directly measured from the images produced
from the STEREO-B vantage point, but its measurements from the
SDO/AIA images are not straightforward. We use for this purpose a
simple geometric model, presented in Figure~\ref{F-model}.

Assuming that the prominence expands in all three dimensions at the
same rate, one might expect its area [$A$] to be proportional to the
squared height of its lower edge [$h^2$]. To find a geometrical
coefficient [$k$] relating the height to the area [$k\,h =
\sqrt{A}$] we represent the sky-plane projection of the crescent
prominence as the overlap of two identical disks [$\mathrm{D}_1$]
and [$\mathrm{D}_2$] of a radius [$R$] (the gray shading in
Figure~\ref{F-model}). The intersections of their outer circles
correspond to the bases of the prominence. Its area is a difference
between the areas of two circular segments, one of which is a
segment of the upper disk $\mathrm{D}_1$ subtended by an angle of
$2\pi - \alpha$, and another is a segment of the lower disk
$\mathrm{D}_2$ subtended by an angle of $\alpha$. The area of a
circular segment subtended by an angle of $\theta$ [radians] is
$R^2(\theta - \sin \theta)/2$, and the difference of the segment
areas is $A = R^2[(2\pi - \alpha) - \sin(2\pi-\alpha)]/2 -
R^2(\alpha-\sin\alpha)/2 = R^2(\pi-\alpha+\sin\alpha)$. The height
of the lower prominence edge is $h = R[1-\cos(\alpha/2)]$, and the
coefficient relating the square root from area to the height is $k =
\sqrt{A}/h = \sqrt{\pi - \alpha + \sin\alpha}\,/\,[1 -
\cos(\alpha/2)]$. When the prominence rises, its legs stretch, and
the circles transform into ellipses. Nevertheless, the coefficient
$k$ determined by the shape of the prominence should not change
considerably within a limited range of height, and correspondence is
expected between the real height of the lower prominence edge [$h$]
and the estimate $\sqrt{A}/k$. The radius $R$ does not stand
explicitly here, being not significant.

\begin{figure} 
     \centerline{
               \includegraphics[width=0.48\textwidth,clip=]
        {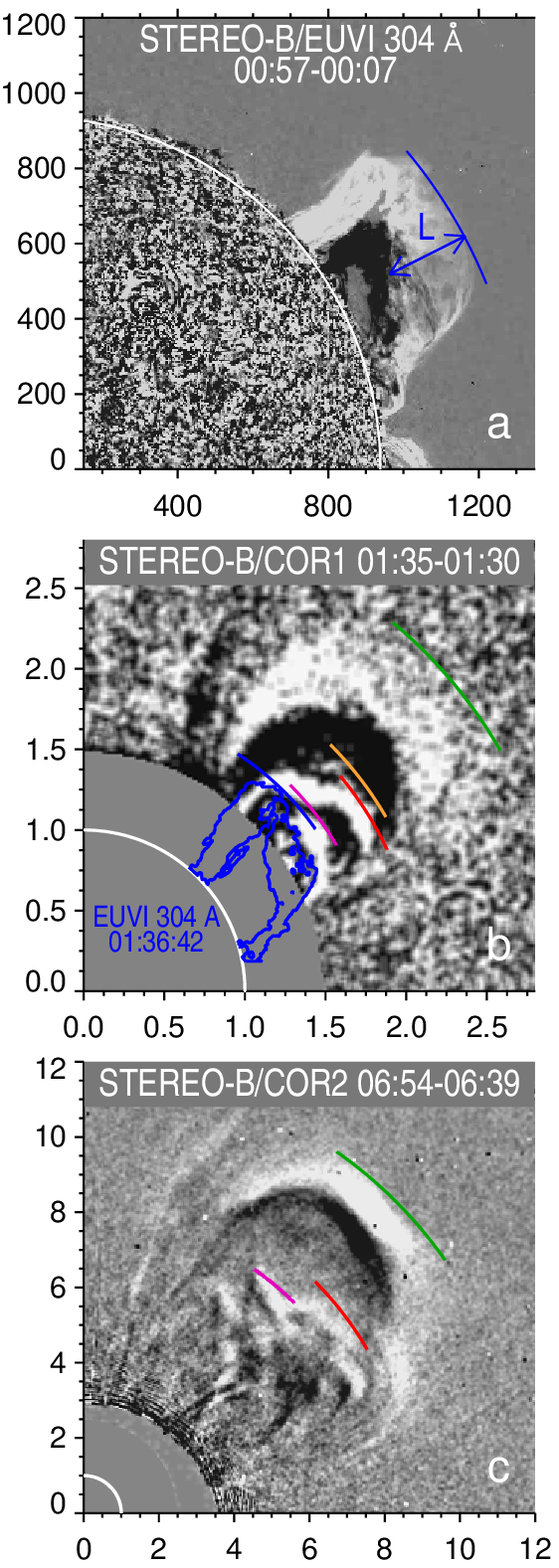}
               \includegraphics[width=0.48\textwidth,clip=]
        {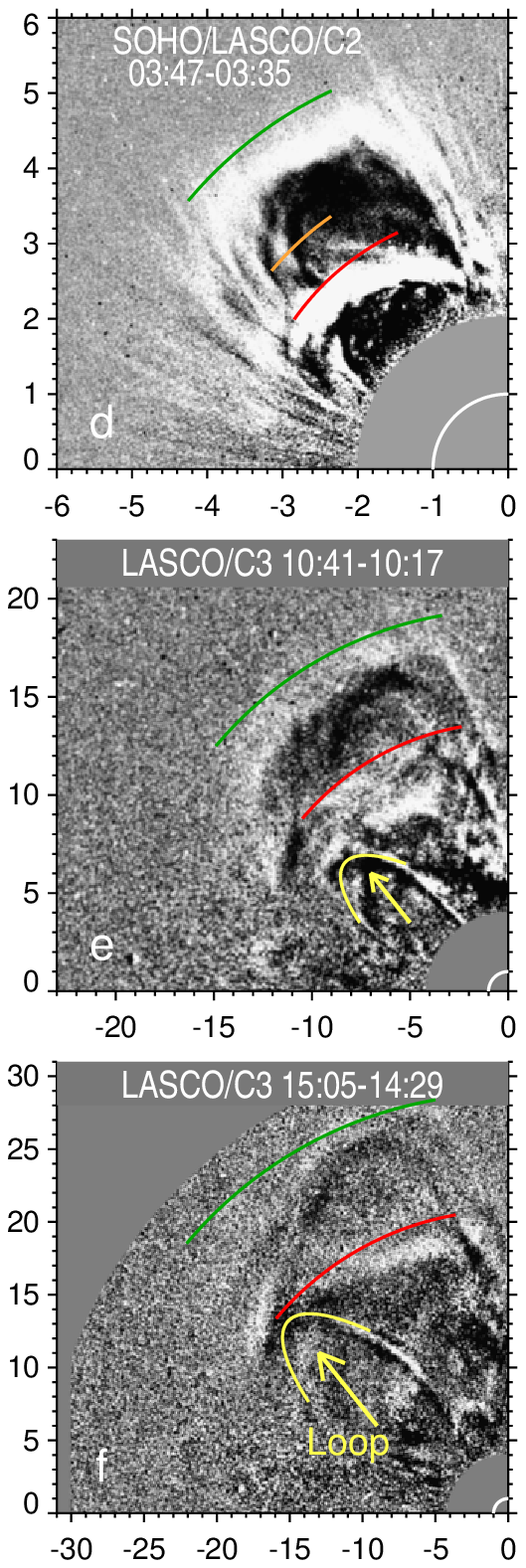}
              }
   \caption{Erupting prominence and CME in STEREO-B (\textit{left column})
and LASCO (\textit{right column}) running-difference images.
(\textbf{a})~Erupting prominence in 304\,\AA\ (EUVI). The
\textit{blue arc} outlines the outer edge of the prominence, whose
position is close to the lower segment of the CME core in panel b.
The axes indicate the distance from solar disk center in arcseconds.
(\textbf{b},~\textbf{c})~CME observed by COR1 (\textbf{b}) and COR2
(\textbf{c}). The \textit{blue contour} in panel b represents the
prominence observed by EUVI in 304\,\AA\ at 01:36.
(\textbf{d}\,--\,\textbf{f})~CME in LASCO-C2 and -C3 images. The
color arcs represent the analytic fit for the prominence
(\textit{blue}), different components of the core (\textit{pink},
\textit{red}, and \textit{orange}), and the leading edge
(\textit{green}). The axes in panels b\,--\,f indicate the distance
from solar disk center in $\mathrm{R}_\odot$.}
   \label{F-stereo_lasco}
   \end{figure}

The height of the lower prominence edge above the limb was
measured from STEREO-B/EUVI 304\,\AA\ images for its middle in the
radial direction (Figure~\ref{F-stereo_lasco}a). The results are
presented by the open circles in Figure~\ref{F-prom_estimates}b.
The triangles show the height estimated from radio absorption. The
crosses represent the estimates based on the prominence area [$A$]
measured from SDO/AIA 304\,\AA\ images. With $k \approx 2$
($\alpha \approx 135^\circ$) the height [$h$] actually measured
from EUVI images and the estimate $\sqrt{A_\mathrm{AIA\,304}}/k$
agree with each other. The decrease of the prominence area after
01:30 could be caused by its decreasing opacity in 304\,\AA\ and
departure from the analyzed region
(Figure~\ref{F-aia304}c).

With the parameters of the erupting prominence found from the model
of radio absorption for different times, its mass can be estimated.
An average electron number density [$n_\mathrm{e}$] was found from
the expression for the optical thickness $\tau \approx 0.2
n_\mathrm{e}^2\,L\,\nu^{-2}T^{-3/2}$, where $\nu$ is a corresponding
frequency (both $\tau$ and $\nu$ are related to a fiducial frequency
of 17\,GHz in our estimates). The geometrical depth of the
prominence [$L$] can be estimated from STEREO-B/EUVI 304\,\AA\
images. When the eruption starts and a negative burst indicates
screening of the Sun, a helical structure of the prominence is
expected to be present (see the \textsf{20130817\_EUVI304.mpg}
movie). Therefore, the cross-section of the prominence was most
likely circular. We measured for each time its width in the radial
direction (Figure~\ref{F-stereo_lasco}a). The mass was estimated as
$m = m_\mathrm{p}n_\mathrm{e}AL$ with $m_\mathrm{p}$ being the
proton mass. The ionization degree of the absorbing material was
assumed to be close to 100\,\%.

The estimated mass is presented in Figure~\ref{F-prom_estimates}c.
The boundaries of the shaded region correspond to the prominence
area estimated from radio absorption and from AIA~304\,\AA\ images.
The triangles represent the average values. The increase of the mass
from $2 \times 10^{15}$\,g to $3.4 \times 10^{15}$\,g reflects the
lift-off and expansion of the prominence. Then the estimated mass
abruptly decreases after 00:30, because the prominence lost opacity
(see Table~\ref{T-model_est}) and exceeded the maximum distance of
$1.186\,\mathrm{R}_\odot$ handled by our model. This decrease
prevented saturation of the plot in Figure~\ref{F-prom_estimates}c,
which would correspond to the approach to the actual mass. To
estimate a probable mass, we fit the increasing part of the plot
with an exponential rise $a[1-\exp \{-(t-t_0)/\tau \}]+b$. The
saturation values [$a+b$] specified in the figure supply a probable
estimate of $\approx 6 \times 10^{15}$\,g. The mass of the
prominence is further discussed in Section~\ref{S-discussion}.

Comparison of the estimates obtained from radio absorption without
imaging data with direct measurements from 304\,\AA\ images confirms
that our model provides realistic parameters for an erupting
prominence (filament), despite its obvious drawback. A reasonable
correspondence between the quantitative parameters of the erupting
prominence estimated from the model and those measured from EUV
images and between their temporal evolutions confirm that the
negative burst was caused in this event exclusively by screening the
quiet-Sun areas, without coverage of any compact microwave source.

\section{Expansion of CME Components}
  \label{S-kinematics}

To study the evolution of the CME associated with the prominence
eruption, in this section we analyze the kinematics of its
structural components. Observations of this CME have the following
advantages: i)~The CME was observed from two vantage points of SOHO
and STEREO-B, ii)~having a rather low speed, the CME was observed in
many images, which makes possible its detailed measurements;
iii)~the structure of the CME core was clearly visible, providing a
rare opportunity to analyze the structural components of the core.

\subsection{Measurements of Kinematics}

For the measurements we used running differences produced from the
images observed by the COR1 and COR2 coronagraphs on STEREO-B and by
the LASCO-C2 and -C3 coronagraphs on SOHO. To co-ordinate the
measurements from SOHO and STEREO-B images, we use the fact that the
visible size of a structure observed from any vantage point is a
linear transformation of its real size. We measured the initial rise
of the prominence and early CME expansion from STEREO-B images,
where they are better visible, and adjusted the scaling factor and
offset for the measurements from SOHO data to match the results
obtained from STEREO-B data. Thus, our measurements are related to
the plane of the sky viewed from STEREO-B. We measured the erupting
prominence, detectable components of the core, and CME front. The
distances measured for the FS [$d$] can be compared with those in
the CME catalog as $d_\mathrm{LASCO} = d_\mathrm{STEREO}/1.05$. We
did not measure the cavity, whose faintness makes it equally
difficult to detect it in non-subtracted images and to distinguish
it from the CME front in running differences.

We used the measurement technique outlined in
Section~\ref{S-introduction}. The distances measured manually were
fitted with an analytic function corresponding to a Gaussian
acceleration pulse, assuming that a huge CME expands gradually. The
measurements made directly from the images were used to estimate the
initial and final velocities. The distances were calculated by
integration of the Gaussian pulse with parameters used as starting
estimates, which were then iteratively refined. If more than one
constant-speed interval show up, than a combination of a few
Gaussian acceleration pulses was used. A final refinement of the
estimated kinematical parameters was made using a movie composed
from the images with a field of view resized according to the
previous-step measurements. An expanding structure of interest
should be static in such a movie. If expansion of a CME is perfectly
self-similar, then all of its structures should be static in a
resized movie. This was not the case in our event. The
\textsf{20130817\_STEREO.mpg} and \textsf{20130817\_LASCO.mpg}
movies were resized according to the measured kinematics of the CME
front, keeping it static. The \textsf{20130817\_STEREO\_core.mpg}
and \textsf{20130817\_LASCO\_core.mpg} movies keep the main part of
the core static.

The errors of the manual distance--time measurements estimated
subjectively are within $\pm 10$\,Mm for the prominence observed
in EUVI 304\,\AA\ images, within $\pm 50$\,Mm for the core in COR1
and C2 images, and within $\pm 200$\,Mm for the core in COR2 and
C3 images. The estimated errors for the FS are within $\pm
100$\,Mm in COR1 and C2 images and within $\pm 300$\,Mm in COR2
and C3 images. These estimates of the errors should be considered
as tentative. The total uncertainties include the errors of the
analytic fit to the distance--time points measured manually. As
mentioned, our ultimate criterion of the measurement quality is a
static state and fixed size of an analyzed structure in a resized
movie.

\subsection{Prominence}

The erupting prominence is visible in EUV and white-light images.
The \textsf{20130817\_EUVI304.mpg} movie presents the prominence in
304\,\AA\ with an upper edge outlined by the blue arc according to
our measurements. These images are not resized. The deviations of
the arc from the prominence edge within $\pm 20$\,Mm characterize
the overall measurement errors. Initially, the prominence was
static. Its lift-off occurred with an acceleration, which reached a
peak of 36\,m\,s$^{-2}$ at 00:59, when its top was located at
$1.42\,\mathrm{R}_\odot$. The acceleration pulse lasted at half
height from 00:28 to 01:32. Conspicuous are the untwisting motion of
the prominence and its complex multi-thread structure. A thin
feature resembling the upper part of a descending bridge is visible
in the movie close to the northern leg between 01:00 and 01:22. Then
this feature disappeared, and the top part of the prominence above
it tended to divide in two parts. This structural change corresponds
to the measured acceleration peak; however, it is not clear so far
if this correspondence is significant. After 01:50, the prominence
top reached a speed of 150\,km\,s$^{-1}$ and became invisible in
304\,\AA. Coronal structures above the rising prominence are not
detectable in EUVI 195\,\AA\ images.

\subsection{CME Components}

Subsequent expansion of the CME is visible in white-light images
produced by the COR1 and COR2 coronagraphs on STEREO-B. The
running-difference movies \textsf{20130817\_STEREO.mpg} and
\textsf{20130817\_STEREO\_core.mpg} show the CME structures with a
high contrast. These images are complex because of subtraction and
the presence of different CME components. They can be identified
with well-known main parts of the CME in the non-subtracted
\textsf{20130817\_cor1\_orig.mpg} movie. The arcs outlining the
middle (red) and north (pink) components of the core and a faint
CME leading edge (green) are only plotted in this movie. The
visible separation of the prominence continued. Its north part
moved faster, apparently disintegrated between 01:36 and 02:15,
stretched, and lost brightness.

The running-difference movies and Figure~\ref{F-stereo_lasco} reveal
more details in the CME structure. A loop-like thick middle
structure outlined by the red arc is visible in
Figure~\ref{F-stereo_lasco}b high above the south part of the
prominence. Being detectable in all white-light images, it was
measured up to the largest distances.

The lowest north segment of the core outlined by the pink arc in
Figures \ref{F-stereo_lasco}b and \ref{F-stereo_lasco}c was
observed by COR1 and COR2 but not by LASCO. The prominence visible
in 304\,\AA\ (blue arcs and contour in Figures
\ref{F-stereo_lasco}a and \ref{F-stereo_lasco}b) was close to this
segment. The different appearance of this core segment in white
light and the prominence in 304\,\AA\ might be the result of the
difference in the spectral ranges, diffraction on the occulting
disk of the coronagraph, and scattered light.

  \begin{figure} 
     \centerline{
               \includegraphics[width=0.48\textwidth,clip=]
        {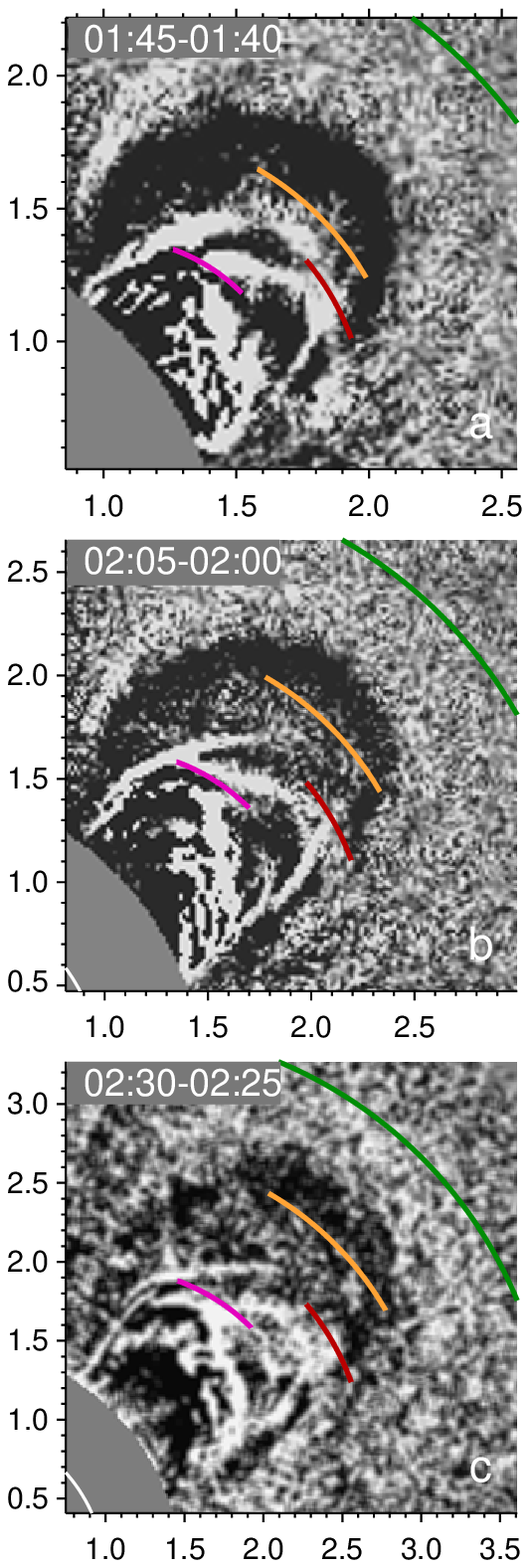}
               \includegraphics[width=0.48\textwidth,clip=]
        {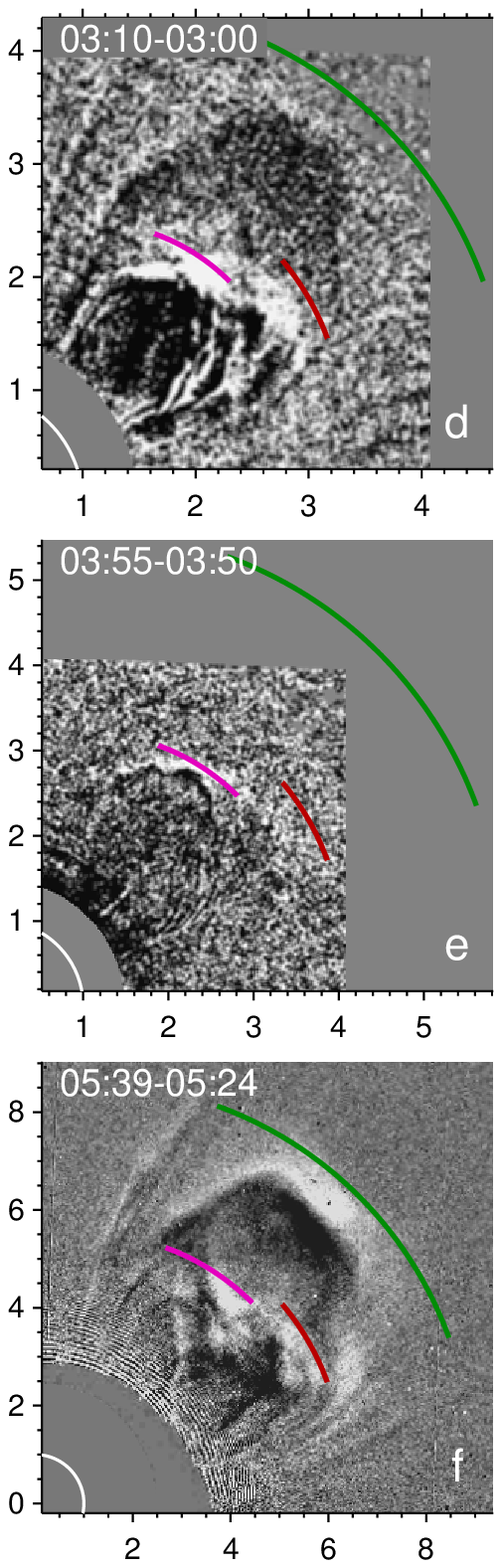}
              }
   \caption{Structural changes of the CME core associated with the
first and second acceleration episodes. The STEREO-B/COR1
(\textbf{a}\,--\,\textbf{e}) and COR2 (\textbf{f}) images are
resized according to the measured kinematics of the middle core
component (\textit{red}). The seemingly different thickness of the
core (especially conspicuous between panels \textbf{e} and
\textbf{f}) is a spurious effect caused by subtracting images
separated by different time intervals.}
   \label{F-cor1_cor2}
   \end{figure}

The fastest loop-like structure is outlined by the orange arc in
Figures \ref{F-stereo_lasco}b\,--\,\ref{F-stereo_lasco}d and
\ref{F-cor1_cor2}a\,--\,\ref{F-cor1_cor2}c, where its evolution is
better visible. Figure~\ref{F-cor1_cor2} presents the images after
acceleration pulses, when the speeds of the accelerated components
considerably increased, making the changes conspicuous. The
fastest structure, whose northern part extended a leg of the
prominence, accelerated earlier and sharper than other parts of
the core. Having appeared after 01:30, this fast structure rapidly
stretched, embraced the whole core, and after 02:00 it disappeared
in the cavity.

\begin{figure} 
   \centerline{\includegraphics[width=\textwidth]
    {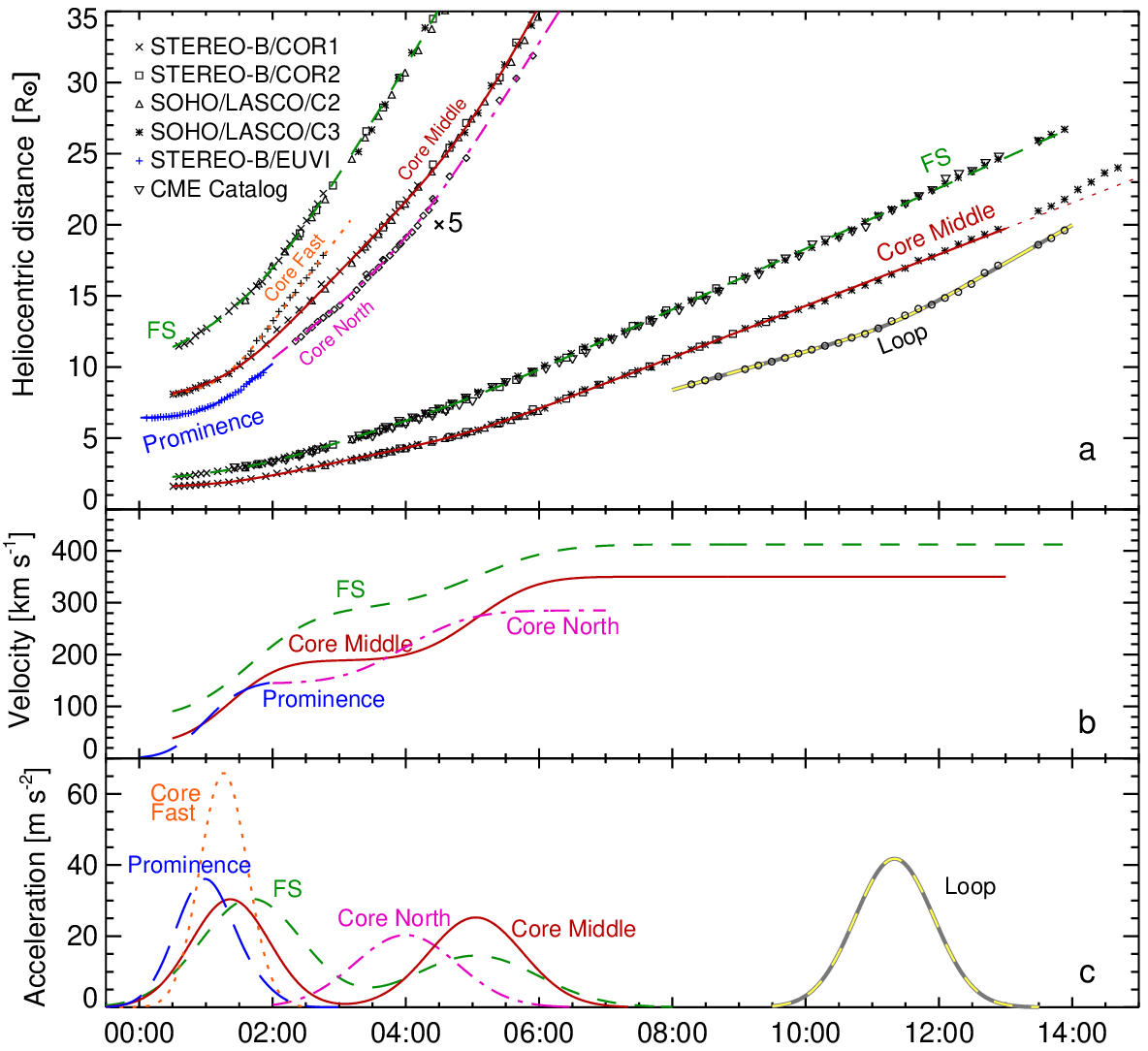}
   }
   \caption{(\textbf{a})~Height--time relation measured from STEREO-B
and SOHO/LASCO images. The \textit{symbols} represent the
heliocentric distances measured for the erupting prominence as well
as different components of the CME core, FS, and the loop. The
measurements from the LASCO data were scaled to match those from the
STEREO-B vantage point. The \textit{down-pointing triangles}
represent the measurements from the CME catalog. The \textit{curves}
represent analytic fit of the measured points. The
\textit{upper-left region} shows the initial portions of the plots
magnified by a factor of five. (\textbf{b})~Velocity--time plots for
the prominence, middle part of the core, and FS.
(\textbf{c})~Accelerations of the prominence, FS, core components,
and the loop. The latest parts of some plots are shown by
\textit{broken lines} to indicate their increased uncertainties.}
   \label{F-kinematics_lasco_stereo}
\end{figure}

The kinematical plots for the core segments and the FS in
Figure~\ref{F-kinematics_lasco_stereo} show that they underwent,
at least, two acceleration episodes. The main parameters estimated
for the CME components are listed in Table~\ref{T-kinem}, which
presents for each acceleration episode the time of the
acceleration peak and the distance of a corresponding structure
from the solar disk center.

\begin{table} 
 \caption{Kinematical parameters of the CME structural components.}
 \label{T-kinem}
 \begin{tabular}{lcrccccc}

 \hline

\multicolumn{1}{c}{CME} & Initial & \multicolumn{6}{c}{Acceleration episode}  \\

\multicolumn{1}{c}{component} & speed & \multicolumn{2}{c}{1} & \multicolumn{2}{c}{2} & \multicolumn{2}{c}{3} \\

 & [km\,s\,$^{-1}$] & \multicolumn{1}{c}{$T_\mathrm{peak}$} & \multicolumn{1}{c}{$r_\mathrm{peak}$} & \multicolumn{1}{c}{$T_\mathrm{peak}$} & \multicolumn{1}{c}{$r_\mathrm{peak}$} & \multicolumn{1}{c}{$T_\mathrm{peak}$} & \multicolumn{1}{c}{$r_\mathrm{peak}$}  \\

 &  &  & \multicolumn{1}{c}{[R$_\odot$]} &  & \multicolumn{1}{c}{[R$_\odot$]} &  & \multicolumn{1}{c}{[R$_\odot$]}  \\

 \hline

Prominence & 0 & 00:59  &  1.42  &  &  &  & \\

Core: &  \multicolumn{7}{c}{}  \\

\quad Fast & 27 & 01:16  &  1.87  &  &  &  & \\
\quad Middle & 27 & 01:22  &  1.93  & 05:03 & 5.6 & 13:22 & 20.7 \\
\quad North &   &   &   & 04:00 & 3.8 & & \\

Front & 78 & 01:41  &  3.07  & 05:03 & 8.0 &  &  \\

Loop &   &   &    &  &  & 11:20 & 13.2 \\

 \hline

 \end{tabular}

 \end{table}

The prominence eruption and early evolution of the CME exhibit
structural changes associated with the first acceleration episode.
Some segments separated from the core, extended forward, taking
the shape of a simple loop, stretched and disappeared in the
cavity. The temporal succession of the acceleration pulses
suggests an outward-propagating disturbance produced by an
innermost structure, \textit{i.e.} the prominence or its invisible
higher-temperature envelope. The CME frontal structure had the
latest response.

Subsequent evolution of the CME is shown by the
\textsf{20130817\_STEREO\_core.mpg} movie and Figures
\ref{F-cor1_cor2}c\,--\,\ref{F-cor1_cor2}f. All of the images are
resized to keep the middle segment of the core static. The faintly
visible structures below the pink arc outlining the top of the north
segment resemble an expanding arcade. They approached the pink arc
after 02:30 and joined the north segment around 03:30, so that the
core in Figures \ref{F-stereo_lasco}c, \ref{F-stereo_lasco}e, and
\ref{F-cor1_cor2}f consists of a few layers of loop-like structures.
As a result, the north segment accelerated around 04:00 and
``pushed'' the middle segment from below. We measured the second
acceleration pulse to be simultaneous for the middle segment of the
core and FS, but certainly later than for the north segment. Like
the first acceleration episode, the disturbance responsible for the
CME acceleration propagated from its inner structures outward. Note
that between the first and second acceleration episodes,
acceleration of the arcade-like structure occurred, which we did not
measure. The structural transformations described here show that the
CME core in this event continued to form up to a heliocentric
distance of $\gsim 4\,\mathrm{R}_\odot$.

\subsection{Last Acceleration Episode}

According to the CME catalog, this CME possessed an overall
acceleration. Besides the apparently accelerating initial part,
Figure~\ref{F-kinematics_lasco_stereo}a shows that the core
accelerated again at a distance of about $21\,\mathrm{R}_\odot$
after 13:00. The top part of the core became faint, but its lower
bright segment is still clearly visible. Comparison of Figures
\ref{F-stereo_lasco}e and \ref{F-stereo_lasco}f reveals that the
lower segment approached the constant-speed fit of the core top.
Because of the large uncertainties, we have not plotted the third
acceleration pulse for the core in
Figure~\ref{F-kinematics_lasco_stereo}; some of its parameters are
listed in Table~\ref{T-kinem}.

The LASCO-C3 images and corresponding movies show from 08:00 to
14:00 a loop-like structure (``Loop'') outlined by the yellow arc in
Figures \ref{F-stereo_lasco}e and \ref{F-stereo_lasco}f. The
distance--time measurements for this structure are presented by the
circles in Figure~\ref{F-kinematics_lasco_stereo}a, and its fitted
acceleration is shown in Figure~\ref{F-kinematics_lasco_stereo}c by
the dashed-yellow curve. The loop accelerated about two hours
earlier than the core, approached it, and pushed the left (in the
plane of the sky) edge of its lower segment. This interaction
resulted in stretch of this edge of the core and FS. Moreover,
acceleration of the CME front is indicated by its position relative
to the green fitting arc corresponding to a constant speed after
10:00.

Finally we note that the distance--time measurements of the CME
core and FS could formally be fitted with a single acceleration
pulse each. In this case, the FS acceleration peak of $\approx
21$\,m\,s$^{-2}$ occurred at 02:46, 12 minutes earlier than that
for the core ($\approx 18$\,m\,s$^{-2}$). The half-height duration
of each acceleration pulse was about 3.5\,--\,4 hours. The
corresponding analytic curves fitted the measured points rather
well, systematically deviating from them within limited time
intervals, especially in the initial stage. With this fit, it was
not clear what could accelerate the CME around 02:50.
Considerations of the changes in the CME structure specified the
kinematics and prompted the possible causes of the acceleration
episodes and a realistic scenario. The detailed measurements
changed the apparent causal relation between the core and FS with
respect to the relation suggested by the fit with a single
acceleration pulse.

\section{Discussion}
  \label{S-discussion}

\subsection{Estimates from Radio Absorption}

The ``isolated'' negative burst observed on 17 August 2013 at
several microwave frequencies was exclusively caused by screening of
the quiet Sun's emission by the prominence material, because no
active regions existed in this part of the solar surface. This
situation is the simplest case for the model of radio absorption
used in our analysis. The model allowed us to estimate the area of
the screen absorbing microwaves, which reached $\approx 10\,\%$ of
the solar disk for the deepest radio depression, larger than the
2\,--\,6\,\% estimated for different events with negative bursts
\citep{KuzmenkoGrechnevUralov2009, Grechnev2013}. The temperature of
the prominence material of 9000\,K corresponds to a typical
situation.

Detailed observations of this event by SDO/AIA and STEREO-B/EUVI
from different vantage points allowed us, for the first time, to
compare the temporal variations of the parameters estimated from
radio absorption with those directly measured from the 304\,\AA\
images. Both methods present similar variations with a quantitative
difference within a factor of two. The temporal sequence of the
estimates promises a more realistic evaluation of the prominence
mass. The extrapolated plausible mass of the prominence found in
Section~\ref{S-estimated_parameters} is $\approx 6 \times
10^{15}$\,g. This estimate is related to low-temperature plasma
only, because hotter structures embracing the prominence are most
likely not detectable in microwaves because of their low opacity.

Our result exceeds the masses of quiescent filaments (prominences)
estimated previously in different studies. \cite{Koutchmy2008}
estimated the mass of an eruptive filament of $2.3 \times
10^{15}$\,g from H$\alpha$ and EUV images, while the mass of the
white-light CME core was $4.6 \times 10^{15}$\,g. However, a
higher-temperature prominence-to-corona interface may have a
considerable mass, not being visible in H$\alpha$ images
\citep{AulanierSchmieder2002}. To overcome the difficulties inherent
for the estimates from observations in the H$\alpha$ line,
\cite{Gilbert2005} developed a simpler method to estimate the mass
of a filament from its absorption of EUV emission.
\cite{Gilbert2006} found an average mass of $4.2 \times 10^{14}$\,g
for static quiescent prominences and $9.1 \times 10^{14}$\,g for
eruptive ones; the authors also listed several reasons for
underestimation of the masses. Using multi-spectral data,
\cite{Schwartz2015} estimated the masses of six static quiescent
prominences from $2.9 \times 10^{14}$\,g to $1.7 \times 10^{15}$\,g.
On the other hand, our extrapolated estimate of $\approx 6 \times
10^{15}$\,g is close to a theoretical result obtained by
\cite{Low2003} for the hydromagnetic equilibrium of a quiescent
prominence, which stores energy sufficient to account for the energy
of a typical CME.

The mass of this CME of $3.6 \times 10^{15}$\,g estimated in the
online CME catalog (\citealp{Yashiro2004};
\url{cdaw.gsfc.nasa.gov/CME_list/}) was most likely concentrated in
its low-temperature core. The CME core usually has a considerably
larger mass than FS, which was also the case in our event, as the
\textsf{20130817\_cor1\_orig.mpg} movie indicates. Thus, the mass of
the CME material at coronal temperatures was presumably $\lsim 1
\times 10^{15}$\,g. Draining of low-temperature material from the
erupting prominence back to the solar surface considerably reduced
its mass and obviously increased the resulting force that drove its
lift-off (see, \textit{e.g.}, \citealp{SchmahlHildner1977,
GopalHanaoka1998, Low2003}). However, unlike the expectations of
these authors, most of the CME mass in the 17 August 2013 event was
supplied by the erupting prominence, while the contribution from its
environment was minor.

\subsection{Causal Relations between CME Structures}

The CME in question was a typical gradually developing
non-flare-related CME. Such CMEs are generally characterized by a
weak ($< 100$\,m\,s$^{-2}$), long-lasting acceleration occurring
in the inner and outer corona \citep{MacQueenFisher1983,
Sheeley1999, Srivastava2000, Zhang2004}. The acceleration pulses
measured for different CME components were comparable with each
other in magnitude and lasted one to two hours at half-height.

The earliest acceleration pulse was measured for the erupting
prominence. Its higher-temperature extension, invisible in 304\,\AA,
corresponded kinematically to the north component of the CME core.
No CME feature exhibited any preceding activity. There is no
indication of anything that could pull the prominence up. Most
likely, nothing but the prominence was a direct driver of the CME.

As the observations show, the acceleration episodes revealed were
associated with the changes in the inner CME structures. The first
acceleration of the core was induced by the prominence eruption.
Then, the fastest core segment accelerated, stretched, and
disappeared in the cavity (the brightness of an expanding CME
structure decreases as the increase in its length squared). Its
acceleration occurred earlier and sharper than that of the middle
segment and FS.

The second acceleration of the middle-core segment and FS was
induced by the north core component, which accelerated one hour
before. In turn, its acceleration was probably caused by the
combination of two loop-like segments visible below it in
STEREO/COR1 movies between 02:10 and 02:50. As \cite{Uralov2002}
showed, the combination of two prominence segments sharply
increases the total twist and, correspondingly, the propelling
force.

The frontal structure accelerated later than the core with a delay
within 25 minutes. The outer edge of the CME appears to be quietly
expanding in all images. No changes in the shape of FS are
visible, which could cause the changes in the core observed.
Moreover, our resized STEREO and LASCO movies demonstrate that the
relative distance between the core and FS progressively decreases,
\textit{i.e.} the core approaches FS. This behavior is not
expected, if the core had been passive, while FS certainly did not
decelerate.

The observations indicate that all changes in the kinematics and
structure of the CME were caused by the processes in its interior
rather than in outer structures. The most active behavior was
exhibited by the erupting prominence (core), while the FS was
forced to expand by an action from inside.

\subsection{Magnetic Field in the CME Cavity}

The temporal sequence of the acceleration pulses of different CME
components reflects an outward-propagating disturbance generated
by internal structures of the core. Most likely, this disturbance
propagated with a fast-mode speed [$V_\mathrm{fast}$]. Using our
measurements, we try estimating magnetic parameters of the CME.

The observed propagation velocity of a fast-mode disturbance
[$V_\mathrm{obs}$] in a moving medium is a sum of the fast-mode
speed and the velocity of the medium. This velocity increases toward
the CME leading edge (depending linearly on the distance for a
perfectly self-similar expansion). For simplicity, we have
subtracted a midway velocity [$V_\mathrm{m}$] between the source and
target, \textit{i.e.} $V_\mathrm{fast} = V_\mathrm{obs} -
V_\mathrm{m}$.

The disturbance propagated in the CME outward nearly perpendicular
to its magnetic field; thus, $V_\mathrm{fast} \approx
\left(V_\mathrm{A}^2 + V_\mathrm{S}^2\right)^{1/2}$ with
$V_\mathrm{A} = B/\sqrt{4 \pi \rho}$ being the Alfv{\'e}n speed,
[$B$] magnetic-field strength, [$\rho$] density, and $V_\mathrm{S}$
the sound speed. If the CME expansion were omnidirectional, then its
parameters change with the increase of the size [$r$] as $B \propto
r^{-2}$ because of magnetic-flux conservation and $\rho \propto
r^{-3}$; hence, $V_\mathrm{A} = V_{\mathrm{A}0}\left( r/r_0
\right)^{-1/2}$, where $V_{\mathrm{A}0}$ and $r_0$ are related to
the initial position of the CME structures near the solar surface.
We assume their temperatures to be within a range of
0.5\,--\,2.5\,MK corresponding to $V_\mathrm{S} = 105 -
235$\,km\,s$^{-1}$.

\begin{figure} 
   \centerline{\includegraphics[width=0.7\textwidth]
    {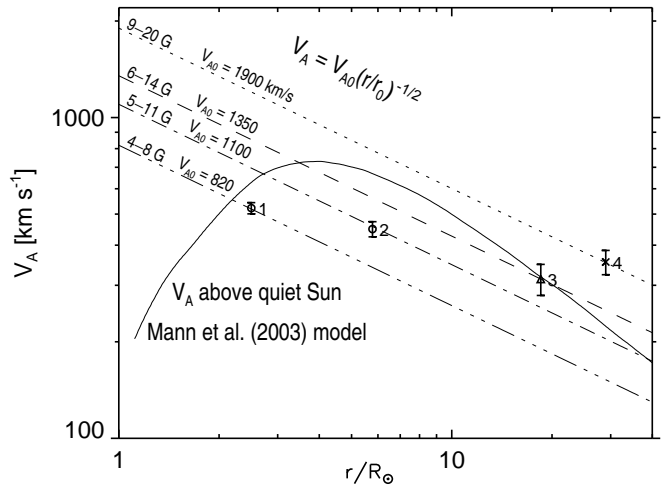}
   }
   \caption{Alfv{\'en} speed in the CME estimated for
four phases of its expansion (\textit{symbols} with \textit{bars})
in comparison with its dependence \textit{vs.} distance expected for
the omnidirectional CME expansion (\textit{broken lines}) and the
model by \cite{Mann2003} for the Alfv{\'en} speed distribution above
the quiet Sun (\textit{solid curve}). The corresponding near-surface
magnetic-field strengths are indicated at the origins of
\textit{slanted broken lines}.}
   \label{F-alfven_speed}
\end{figure}

The Alfv{\'e}n speed in the CME that is estimated in this way for
four expansion episodes is shown by symbols in
Figure~\ref{F-alfven_speed}. They represent propagation from the
middle core segment to FS in acceleration episode 1 (point 1), from
the north core segment to FS in episode 2 (point 2), from the loop
to the middle core segment in episode 3 (point 3), and from the loop
to FS (point 4). The acceleration time of the FS for point 4 was
estimated approximately, without accurate measurements, because of
the poor FS visibility. All measured propagation velocities are of
the same order: $V_\mathrm{obs} = 700-800$\,km\,s$^{-1}$. The bars
correspond to the temperature range of 0.5\,--\,2.5\,MK. The
slanted-broken lines crossing the four measured points represent the
$V_\mathrm{A} = V_{\mathrm{A}0}\left( r/r_0 \right)^{-1/2}$
dependence.

Points 1, 2, and 4 in Figure~\ref{F-alfven_speed} correspond to the
CME cavity, while point 3 corresponds to a rarefied volume below the
core. The number density of the coronal plasma in a prominence
cavity near the solar surface is probably within a range of $(1-5)
\times 10^{8}$\,cm$^{-3}$ (which also seems to apply to the
back-extrapolated volume below the core). The near-surface
magnetic-field strengths corresponding to this density range are
listed near the origins of the slanted broken lines. For comparison,
the solid curve represents the model Alfv{\'en} speed distribution
above the quiet Sun \citep{Mann2003}. With a low plasma density in
the cavity, the magnetic fields corresponding to points 1, 2, and 3
do not seem to be strong relative to the environment.

Specifically, the back-extrapolated Alfv{\'e}n speed at point 1
corresponds to 4\,--\,8\,G, which is somewhat weaker than that
expected in a quiescent prominence. However, as the CME expanded,
the magnetic field in its cavity exhibited a relative
strengthening. Point 3 representing the volume below the core also
corresponds to this tendency. This process indicates that the
formation of the CME magnetic structure, including the cavity, was
still in progress during the CME expansion in the outer corona.

\subsection{Formation of CME Structures}

Magnetic-flux ropes (MFR) are believed to be the main active
structures of CMEs, in accordance with a scenario initially proposed
by \cite{Hirayama1974}. Due to numerous observational studies and
theoretical considerations, some stages in the development of an MFR
in a typical CME appear to become clearer.

A probable progenitor of an MFR is a prominence (filament) or a
similar sheared structure, whose temperature is higher. The
prominence together with its cavity resembles a multitude of
MFR-like sections, each of which is connected to the solar surface
separately, while their axes are aligned  parallel to the neutral
line \citep{Gibson2015, Grechnev2015}. Descending prominence threads
are strongly sheared. If for some reason reconnection between the
descending threads of adjacent MFR-like sections occurs, then the
sections join, and they share a combined magnetic field, while the
site of their contact detaches from the photosphere
\citep{InhesterBirnHesse1992}. The poloidal flux in the prominence
increases, and its transformation into an MFR starts. The propelling
Lorentz force grows \citep{Chen1989, Chen1996}. The helical
structure of the prominence becomes pronounced.

As the reconnection process progresses, at some level the prominence
loses equilibrium, and a magnetohydrodynamic (MHD) instability of an
increasing current in it develops, triggering also the
standard-model reconnection in the embracing arcade
\citep{Uralov2002, Grechnev2015, Grechnev2016}. The prominence
erupts; nevertheless, it is unlikely that all of the MFR-like
sections constituting its body have completely combined to form a
single perfect flux rope connected to the photosphere by two ends
only. Separate lateral connections and other residuals of the former
prominence structure are possible.

In fact, the erupting MFR-like structures revealed recently in a few
flare-related events appeared in EUV as complex bundles of hot loops
\citep{Cheng2011, Cheng2013, Grechnev2016}. Many white-light CMEs
also possess complex configurations. On the other hand, some other
CMEs look simpler. Furthermore, \textit{in-situ} measurements often
show nearly perfect structures of interplanetary magnetic clouds
(\textit{e.g.} \citealp{Lui2011}). These facts suggest that the MFR
formation processes possibly continue during the CME expansion, and
the configurations of erupting structures observed near the Sun,
white-light CMEs, and interplanetary magnetic clouds might be
considerably different.

The development of the 17 August 2013 CME appears to confirm this
assumption. The structure of the CME core had not established until,
at least, $4\,\mathrm{R}_\odot$. One of the observed episodes of its
formation is associated with a rise of an arcade-like structure
joining the core from below, which resulted in the second
acceleration pulse. Note that in a free self-similar expansion the
distance between different CME features only increases, while the
ratio of their sizes remains constant.

The leading part of the core also underwent dynamic changes. Some of
its structures straightened, stretched and disappeared in the
cavity. Straightening a twisted structure decreased its brightness
and magnetic-field strength, while the magnetic field became more
uniform and strengthened in the cavity. This process confirmed by
Figure~\ref{F-alfven_speed} indicates that the MFR in the cavity was
probably formed from tangled structures of the core.

While the initial acceleration episode and corresponding
structural transformations constituted a necessary stage creating
the CME, other acceleration episodes revealed in its expansion do
not seem to be crucial milestones of its development. More
probably, the whole evolution of a CME comprised a multitude of
structural changes, which simplified its structure and eventually
transformed it into a more or less perfect flux rope.

A probable progenitor of the CME frontal structure was the coronal
arcade embracing the prominence. While the inner layers of the
arcade are expected to participate in the standard-model
reconnection, its outer loops were stretched by the erupting
prominence, which compressed them from below. The pileup
constituted the frontal structure. A similar scenario was observed
previously in flare-related eruptions \citep{Cheng2011,
Grechnev2015, Grechnev2016}.

\subsection{CME Expansion}

CMEs are affected by several forces, whose roles at different stages
have not yet been established with certainty. These are the
outward-directed magnetic pressure and Lorentz force, the thermal
pressure force, the inward-directed magnetic tension due to the
toroidal field, gravity forces, and aerodynamic drag from the solar
wind (see, \textit{e.g.}, \citealp{Low1982, Chen1989, Chen1996,
ChenKrall2003}). Most studies relate the main propelling force
responsible for the initial lift-off of the majority of CMEs to the
Lorentz force.

The story following the termination of the MHD instability, which
determines the impulsive acceleration stage, seems to be ambiguous.
If within some range of distances the magnetic forces, plasma
pressure, and gravity exceed the drag force, then the CME expands
freely in the self-similar regime \citep{Low1982, Uralov2005}. Such
expansion of many CMEs is well known from observations. Eventually,
drag is expected to become important; indeed, \cite{Gopalswamy2000}
found that slow CMEs were accelerated and fast CMEs were
decelerated, so that the speeds of interplanetary CMEs (ICMEs) at
1\,AU tend to approach the solar wind speed. It is not clear when
drag becomes significant. \cite{Chen1989, Chen1996} and
\cite{ChenKrall2003} consider it to be important even in the inner
corona. Slow CMEs were often considered to be accelerated by the
solar wind; however, the analysis of seven such events by
\cite{Sachdeva2014} shows that aerodynamic drag alone cannot account
for their acceleration. According to \cite{Vrsnak2006} and
\cite{Temmer2011}, drag dominates at distances $>
15-20\,\mathrm{R}_\odot$. However, a huge ICME, which hit Earth on
29 October 2003 with a speed of about 1900\,km\,s$^{-1}$,
surprisingly did not exhibit an expected deceleration
(\citealp{Grechnev2014mc}, Section~3.1). \cite{Rollett2014}
demonstrated that propagation of a CME can be affected by variable
conditions in its way depending on preceding CMEs. These
circumstances show that the role of aerodynamic drag is complex and
needs better understanding.

\begin{figure} 
   \centerline{\includegraphics[width=0.5\textwidth]
    {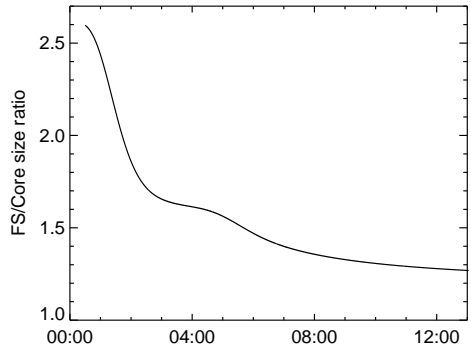}
   }
   \caption{Temporal variations in the ratio between the instantaneous size
of the frontal structure and that of the core relative to the
expansion center. The ratio was calculated from the distance--time
plots in Figure~\ref{F-kinematics_lasco_stereo}a.}
   \label{F-fs_core_ratio}
\end{figure}

The expansion of the 17 August 2013 CME seems to be somewhat
atypical. Unlike many other CMEs, its self-similar regime was not
established even in the outer corona. This fact is obvious from the
resized movies, which show a systematic decrease of the relative
distance between the core and FS. Figure~\ref{F-fs_core_ratio}
quantifies the relation between the sizes of the FS and core by
13:00, excluding the outermost acceleration episode, which we did
not measure. According to \cite{Uralov2005}, the self-similar
expansion is generally characterized by acceleration, which does not
increase in the absolute value. This was not the case in the second
and third acceleration episodes. Furthermore, the distances between
all CME structures increase in the self-similar regime, whereas the
approach of the lower arcade-like structure to the core during the
second acceleration episode presents an opposite process.

The flux-rope model predicts a peak acceleration at a distance [$Z$]
within a range of $S/2 < Z < 3S/2$, where $S$ is the distance
between the bases of the flux rope \citep{ChenKrall2003}. The actual
distance between the bases of the erupting prominence was $S \approx
0.5\,\mathrm{R}_\odot$. However, the distances of
$3.8-8.0\,\mathrm{R}_\odot$ where different CME components underwent
the second acceleration episode (Table~\ref{T-kinem}) were much
larger than the model prediction.

The particularities of the CME expansion were unlikely to have been
related to solar wind, whose largest influence is expected for the
FS, whereas all of the changes started deep inside the CME. The
difference between the speeds of the FS and solar wind was
insignificant, especially in the third acceleration episode.

This latest episode undergone by the core around 13:20 occurred at a
distance of about $21\,\mathrm{R}_\odot$, two hours after
acceleration of the loop, which began ``pushing'' the left edge of
its lower segment. The cause of the acceleration of the loop is not
known. The \textsf{20130817\_LASCO.mpg} and
\textsf{20130817\_LASCO\_core.mpg} movies sometimes show an ongoing
rise of material from behind the occulting disk of the C3
coronagraph, while the source of this trailing material is
uncertain; no associated surface activity is detectable. In any
case, the last acceleration episode demonstrates that the CME
expansion was determined by magnetic forces and plasma pressure in
its inner structures rather than outer drag. Note that no CME
occurred in this sector at least one day before, so that coronal
conditions were unlikely disturbed considerably.

Thus, the particularities found in the expansion of this CME are
not accounted for by known models. Probably, we are dealing here
with an unknown intermediate stage of the CME development between
the initial impulsive acceleration and free self-similar
expansion. This stage was revealed due to the huge size of the
quiescent erupting prominence determining its long-lasting gradual
acceleration and advantages of the resized movies, which made
kinematical and structural changes of the CME conspicuous.
Speculating from the size scale, one might expect that this
``in-flight'' formation stage occurs at much shorter distances for
flare-related CMEs. Here this stage encompassed the first and
second acceleration episodes up to about $10\,\mathrm{R}_\odot$
with an initial size of the erupting prominence of about
$0.5\,\mathrm{R}_\odot$. For a flare-related eruption of a
prominence, whose initial size is less by a factor of 10\,--\,20,
the corresponding CME formation stage is expected to occur behind
the occulting disc of LASCO-C2. This explains why this stage was
not detected previously. As the third acceleration episode
suggests, the CME formation can continue at large distances.
Therefore, the structures of the eruptions observed in EUV, the
white-light CMEs, and ICMEs can have considerable differences.

The aerodynamic drag was unlikely to have been important for this
CME at all, because its speed was close to that of solar wind. On
the other hand, it can be important all of the time for some slow
CMEs, which accelerate very gradually, especially if no associated
surface activity is observed \citep{MacQueenFisher1983,
Robbrecht2009, WangZhangShen2009}.

\section{Summary}
 \label{S-conclusion}

Our analysis of the 17 August 2013 eruptive event was inspired by
a rare ``isolated'' negative burst without any impulsive burst.
Unlike many other negative bursts, its appearance at several
microwave frequencies was exclusively caused by absorption of
the quiet Sun's emission in cool plasma of the erupting prominence,
which screened a considerable part of the Sun. Using the
multi-frequency total-flux data and detailed observations in
304\,\AA\ from two different vantage points of SDO and STEREO-B,
it has become possible for the first time to follow and compare
the temporal variations of geometrical parameters of the erupting
prominence estimated by means of different methods. In particular,
model estimates of the area and height of the prominence from
radio absorption and their direct measurements from EUV images
present similar variations with a quantitative difference within a
factor of two.

The bulk of the prominence material had an average temperature of
9000\,K and a probable total mass of about $6 \times 10^{15}$\,g at
the onset of the eruption. During the lift-off, a part of the cool
prominence material drained back to the solar surface; nevertheless,
the prominence supplied most of the CME mass ($3.6 \times
10^{15}$\,g), while its coronal-temperature part did not exceed
$10^{15}$\,g.

To study the CME lift-off and subsequent expansion, we analyzed
kinematics of its components along with transformations in its
structure. The direct distance--time measurements were used as
starting estimates, which were fit with an analytic function. The
results were refined by means of the movies, whose field of view
continuously increases according to the measured distance--time fit.
The resized movies facilitate verifying the measurements and
revealing any changes in the CME shape and structure. Relative to
the approach based on differentiation of the measurements, this
method is less sensitive to the irregular appearance of CME
structures in the images and produces lesser spurious effects, but
it requires much more effort and time. The results show the
following.

\begin{enumerate}

 \item The main driver of the CME initiation was the prominence.
It was most active and accelerated earlier than any other observed
structures. Then the erupted prominence became the CME core in
agreement with a traditional view.

 \item The core was still active in the course of subsequent CME
expansion. The kinematical and structural changes started in the core
and propagated outward. The frontal structure responded with a
considerable delay.

 \item The CME structures continued to form during its
expansion. The core formed up to $4\,\mathrm{R}_\odot$ with
participation of structures rising behind it.

 \item The cavity also evolved during the CME expansion. Some
structures separated from the core, stretched, and occupied the
cavity. This process possibly transformed tangled structures of
the core into a simpler flux rope, which grew and filled the
cavity.

 \item Most likely, the CME frontal structure formed from coronal
loops embracing the erupting prominence stretched by its
expansion. Throughout the initiation and expansion of the CME
observed, the frontal structure was passive.

\end{enumerate}

Atypically, the self-similar regime of the CME expansion had not
established even up to about $30\,\mathrm{R}_\odot$, while the
role of aerodynamic drag was insignificant. This behavior of the
CME is explained by the phenomena listed. Due to the huge size and
gradual acceleration of the prominence, an intermediate in-flight
stage of the CME development between the initial impulsive
acceleration and free expansion was probably observed. This
possibility indicates that the structures, properties, and roles
of different components of a near-surface eruption, CME, and ICME
may change during their overall history.

\begin{acks}

We thank A.M.~Uralov for recommendations and discussions. We are
indebted to the anonymous reviewer for useful remarks. We thank the
instrument teams of SDO/AIA, STEREO/SECCHI, and SOHO/LASCO (ESA and
NASA); Nobeyama Radio Polarimeters; USAF RSTN Network; and the LASCO
CME catalog generated and maintained at the CDAW Data Center by NASA
and the Catholic University of America in cooperation with the Naval
Research Laboratory. SOHO is a project of international cooperation
between ESA and NASA. The study was supported by the Russian State
Contracts No.~II.16.3.2 and No.~II.16.1.6.

\end{acks}

\section*{Disclosure of Potential Conflicts of Interest} The authors
declare that they have no conflicts of interest.

\end{article}

\end{document}